\documentclass[aps,pra,twocolumn]{revtex4}
\usepackage[dvips]{epsfig}
\usepackage[usenames]{color}
\usepackage{pstricks}
\usepackage{amsmath}
\usepackage{amssymb}
\usepackage{subfigure}
\usepackage{afterpage}
\usepackage{nameref}
\usepackage{array}
\usepackage{multirow}

\begin{document}
\title{Shielding $^2\Sigma$ ultracold dipolar molecular collisions with electric fields}

\author{Goulven Qu{\'e}m{\'e}ner}
\affiliation{Laboratoire Aim{\'e} Cotton, CNRS, Universit{\'e} Paris-Sud, ENS Cachan, Universit{\'e} Paris-Saclay, 91405 Orsay, France} 
\author{John L. Bohn}
\affiliation{JILA, NIST, and Department of Physics, University of Colorado,
Boulder, Colorado 80309-0440, USA}

\date{\today}

\begin{abstract}

The prospects for shielding ultracold, paramagnetic, dipolar molecules from inelastic and chemical collisions are investigated.  Molecules placed in their first rotationally excited states are found to exhibit effective long-range repulsion for applied electric fields above a certain critical value, as previously shown for non-paramagnetic molecules. This repulsion can safely allow the molecules to scatter while reducing the risk of inelastic or chemically reactive collisions.  Several molecular species of $^2\Sigma$ molecules of experimental interest -- RbSr, SrF, BaF, and YO --  are considered, and all are shown to exhibit orders of magnitude suppression in quenching rates in a sufficiently strong laboratory electric field.  It is further shown that, for these molecules described by Hund's coupling case b,  electronic and nuclear spins play the role of spectator with respect to the shielding. 

\end{abstract}


\maketitle
\section{Introduction}

A main direction in the creation and study of ultracold molecules is to control interactions and chemistry~\cite{Krems_PCCP_10_4079_2008,Quemener_CR_112_4949_2012,
Lemeshko_MP_111_1648_2013,Tscherbul_PRL_115_023201_2015}.
One compelling goal of such control would be to prepare reactant species in well-defined, single quantum states in {\it all} degrees of freedom, including individual partial waves of the relative motion.  This would represent the logical endpoint of molecular beam studies, providing results on collision dynamics that are no longer averaged over partial waves, thus yielding more direct and detailed comparisons with theoretical calculations.  Reduction of molecular collisions to either $l=0$ or $l=1$ partial waves has already been achieved in the reaction 2~KRb $\to$ K$_2$ + Rb$_2$~\cite{Ospelkaus_S_327_853_2010}.

Moreover, in the ultracold regime, molecules that possess electric dipole moments are strongly subject to collisional manipulation via electric fields.  Since dipoles can either attract or repel one another, depending on circumstances such as field orientation~\cite{Ticknor_PRA_84_032702_2011,Quemener_arXiv_1508_02289_2015,Frisch_arXiv_1504_04578_2015} and quantum state, electric fields can either enhance or reduce the propensity for the reactant molecules to get close enough to react.  This kind of electric field control of kinematics was also demonstrated in the KRb gas \cite{Ni_N_464_1324_2010}.  

Experimentally, it remains a challenge to produce molecules in the ultracold regime.  Recent progress, notably direct laser cooling, promises molecular samples in the mK temperature range; for example, SrF molecules have very recently been captured in a magneto-optical trap  (MOT)~\cite{Barry_N_512_286_2014}. To proceed to ultracold temperatures would require evaporative cooling, whereby the highest-energy molecules are siphoned off and the remainder come to thermal equilibrium via elastic collisions.  But this procedure comes with a catch: if the molecules are reactive, e.g., by the reaction 2~SrF $\to$ SrF$_2$ + Sr~\cite{Meyer_PRA_83_032714_2011}, then they may be lost to this reaction before coming into thermal equilibrium.  

For this reason, it would be desirable to suppress chemical reactivity even above the ultracold regime, at the 0.1-10 mK temperatures of the MOT.  Luckily, the repulsive dipolar forces between polar molecules can be harnessed for this purpose, as was observed recently in the successful evaporative cooling of OH molecules~\cite{Stuhl_N_492_396_2012}. The example of OH is instructive, as it suggests a general concept.  Namely, each OH molecule possesses a nearly-doubly-degenerate ground state, consisting of a pair of states of opposite parity split by the $\Lambda$-doubling.  Upon bringing two such molecules together, the electric dipole-dipole interaction between molecules mixes these states together, partially polarizing them.  The net result is that the molecules exert forces on one another that are second-order in the dipole-dipole interaction.  There result effective van der Waals interactions, $C_6/R^6$, with coefficients $C_6$ that scale as the fourth power of the molecules' permanent electric dipole moment, and in inverse proportion to the $\Lambda$-doublet splitting $\Delta$~\cite{Avdeenkov_PRA_66_052718_2002}. Moreover, the coefficient $C_6$ is negative for molecules in the lower $\Lambda$-doublet state, in which case the molecules attract one another; and $C_6$ is positive for molecules in the upper $\Lambda$-doublet state, where the molecules therefore repel each other. This circumstance is what makes evaporative cooling possible, although it is complicated by details of the arrangement of electric and magnetic fields within the trap~\cite{Quemener_PRA_88_012706_2013}. 

There remain various categories of ultracold molecules that do not possess $\Lambda$ doublets, however. Notable among these are alkali dimers with $^1\Sigma$ ground states, as well as molecules of $^2\Sigma$ symmetry, such as  SrF~\cite{Barry_N_512_286_2014,McCarron_NJP_17_035014_2015}, RbSr~\cite{Pasquiou_PRA_88_023601_2013}, YO~\cite{Hummon_PRL_110_143001_2013, Collopy_NJP_17_055008_2015,Yeo_PRL_114_223003_2015}, all of current experimental interest. These too may suffer chemical reactions, for example, 2~RbSr $\to$ Rb$_2$ + Sr$_2$~\cite{Zuchowski_PRL_105_153201_2010,Guerout_PRA_82_042508_2010,
Stein_EPJD_57_171_2010,Seto_JCP_113_3067_2000}.
It is to the latter class of  molecules that we direct our attention in this paper.  Such molecules do not possess a $\Lambda$ doublet; rather the states of opposite parity consist of rotational states.  In previous work of Avdeenkov et al.~\cite{Avdeenkov_PRA_73_022707_2006}, refined by Wang et al.~\cite{Wang_NJP_17_035015_2015}, it was found that the shielding mechanism can still be effective for $^1\Sigma$ molecules.  Namely, a pair of molecules in the rotationally excited state $n=1$, at a suitable value of a static electric field, finds their interaction potential raised from below by rotational states of opposite parity, thus generating a shielding potential. In the present article, we show that this is also the case for $^2\Sigma$ molecules.  The presence of electronic and nuclear spin does not disrupt the fundamental shielding mechanism. We note that a similar kind of mechanism can be engineered in the presence of microwave radiation, resulting in suppression of loss~\cite{Gorshkov_PRL_101_073201_2008}.

In Section II, we briefly recall the characteristic properties of the $^2 \Sigma$ molecules and the formalism of their collisional dynamics. In Section III, we present in detail numerical results for RbSr + RbSr collisions. The results demonstrate that the degrees of freedom corresponding to fine and hyperfine structure remain essentially spectators during inelastic and reactive scattering nearby the shielding electric field.  
In Section IV, armed with this insight, we investigate the role of the collision energy for the RbSr system. We also apply the same formalism for collisions involving other experimentally relevant molecules: SrF, BaF, and YO. In all cases, calculations show that the ratio of elastic to quenching collisions can far exceed 100 for realistic collision energies and above a threshold electric field. These results bode well for the prospects of evaporative cooling in these species.  We conclude in Section V.

\section{Theoretical formalism}

As in many ultracold scattering calculations, the main feature of the model is that long-range physics of the molecules is treated in considerable detail, including long-range forces such as dipole and van der Waals forces. Short-range forces, as described by electronic potential energy surfaces are treated more schematically, and chemical reaction is included by means of an absorbing boundary condition. This approach focuses on the physics of the long-range interaction's ability to shield the molecules from reaching the reaction zone at all.

To this end, we consider $^2\Sigma$ molecules in their ground electronic and vibrational state.
The molecules have a fine and hyperfine structure since they 
possess non-zero electronic and nuclear spin. 
The Hamiltonian for a molecule at zero magnetic field is given by:
\begin{eqnarray}
H_\text{mol} = H_\text{rot} + V_{ns} + V_{si} + V_S. 
\label{HAM}
\end{eqnarray}
The first term $H_\text{rot}$ represents the rotational structure of the molecule, 
the second term  $V_{ns} = \gamma_{ns} \ \vec{n}.\vec{s}$ 
represents the rotation-electronic spin coupling,
the third term $V_{si} = b_{si} \ \vec{s}.\vec{i}$ the electronic spin-nuclear spin coupling
and the last term represents the Stark effect $V_S = - \vec{d}.\vec{E}$.
The electric field $\vec{E}$ is chosen as the quantization axis.
We start from a total uncoupled basis set $|n,m_n,s,m_s,i,m_i\rangle$ in ket notation for the molecule
representing respectively the rotation, electronic and nuclear spin quantum numbers.
In this basis set, the matrix elements of the Hamiltonian are given by
\begin{eqnarray}
\langle n,m_n,s,m_s,i,m_i | H_\text{rot} | n',m_n',s,m_s',i,m_i'\rangle = \nonumber \\ 
B_{rot} \ n \,(n+1) \times \delta_{m_n,m_n'} \, \delta_{m_s,m_s'} \, \delta_{m_i,m_i'} ,
\end{eqnarray} 
\begin{multline}
\langle n,m_n,s,m_s,i,m_i | V_{ns} | n',m_n',s,m_s',i,m_i'\rangle =  \\ 
\gamma_{ns} \ m_n \, m_s 
\times \delta_{n,n'} \, \delta_{m_n,m_n'} \, \delta_{m_s,m_s'} \, \delta_{m_i,m_i'}  \\
+ \frac{\gamma_{ns}}{2} \ \left( n \,(n+1) - m_n' \,(m_n' \pm 1)  \right)^{1/2} \\
\times \left( s \,(s+1) - m_s' \,(m_s' \mp 1)  \right)^{1/2} \\ 
\times \delta_{n,n'} \, \delta_{m_n,m_n'\pm1} \, \delta_{m_s,m_s'\mp1} \, \delta_{m_i,m_i'},
\end{multline} 
\begin{multline}
\langle n,m_n,s,m_s,i,m_i | V_{si} | n',m_n',s,m_s',i,m_i'\rangle =  \\ 
\gamma_{si} \ m_s \, m_i 
\times \delta_{n,n'} \, \delta_{m_n,m_n'} \, \delta_{m_s,m_s'} \, \delta_{m_i,m_i'}  \\
+ \frac{\gamma_{si}}{2} \ \left( s \,(s+1) - m_s' \,(m_s' \pm 1)  \right)^{1/2} \\
\times \left( i \,(i+1) - m_i' \,(m_i' \mp 1)  \right)^{1/2} \\ 
\times \delta_{n,n'} \, \delta_{m_n,m_n'} \, \delta_{m_s,m_s'\pm1} \, \delta_{m_i,m_i'\mp1} ,
\end{multline} 
and
\begin{eqnarray}
\langle n,m_n,s,m_s,i,m_i | V_{S} | n',m_n',s,m_s',i,m_i'\rangle = \nonumber \\ 
 - d \, E  \  \, (-1)^{m_n} \, \sqrt{2n+1} \, \sqrt{2n'+1} \nonumber \\
\times  \left( \begin{array}{ccc} n & 1 & n' \\ 0 & 0 & 0 \end{array} \right)
\, \left( \begin{array}{ccc} n & 1 & n' \\ -m_{n} & 0 & m_{n}' \end{array}  \right) \nonumber \\
\times \delta_{m_n,m_n'} \, \delta_{m_s,m_s'} \, \delta_{m_i,m_i'}.
\end{eqnarray}
The total molecular angular momentum projection quantum number 
$m_{f_i} = m_{n_i} + m_{s_i} + m_{i_i}$ for the molecule $i=1,2$ 
is conserved in an electric field.
The different parameters of the molecules are reported in Table \ref{TAB1}.

\begin{table} [t]
\setlength{\extrarowheight}{4pt}
\begin{tabular}{|c|c|c|c|c|}
  \hline
   molecule & $^{87}$Rb$^{84}$Sr & $^{84}$Sr$^{19}$F & $^{138}$Ba$^{19}$F & $^{89}$Y$^{16}$O \\
   \hline
   Ref. & \cite{Guerout_PRA_82_042508_2010,Zuchowski_PRL_105_153201_2010,Zuchowski_2013} & \cite{Steimle_JMS_68_134_1977,Childs_JMS_87_522_1981,Ernst_CPL_113_351_1985} & \cite{Ryzlewicz_CP_51_329_1980,Ernst_JCP_84_4769_1986} & \cite{Bernard_AJSS_52_443_1983,Childs_JCP_88_598_1988,Suenram_JCP_92_4724_1990} \\
  \hline
  $B_{rot}$ (MHz) & 539.6 & 7517.3 & 6471.0 & 11633.6\\
  $\gamma_{ns}$ (MHz) & 20 & 74.79485 & 80.923 & -9.2254  \\
  $b_{si}$ (MHz) & 2786 & 97.0834 & 63.509 & -762.976 \\
  $d$ (D) & 1.54 & 3.47 & 3.17 & 4.52 \\
   \hline
  $s$  & 1/2 & 1/2 & 1/2 & 1/2 \\
  $i$  & 3/2 & 1/2 & 1/2 & 1/2 \\
  \hline
\end{tabular}
\caption{Parameters for the $^2 \Sigma$ molecules of RbSr, SrF, BaF and YO.}
\label{TAB1}
\end{table}

\begin{figure} [h]
\begin{center}
\includegraphics*[width=6.7cm,keepaspectratio=true,angle=-90]{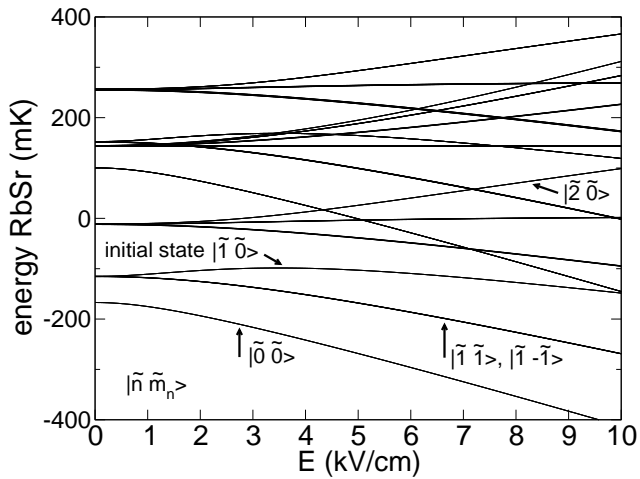} 
\end{center}
\caption{Energy spectrum of a single RbSr molecule in an electric field.
The dressed state are noted $|\tilde{n},\tilde{m}_n\rangle$, see text.
The initial state considered for scattering is $|\tilde{1},\tilde{0}\rangle$.}
\label{NRG-1PLE-FIG}
\end{figure}

\begin{figure} [h]
\begin{center}
\includegraphics*[width=6.5cm,keepaspectratio=true,angle=-90]{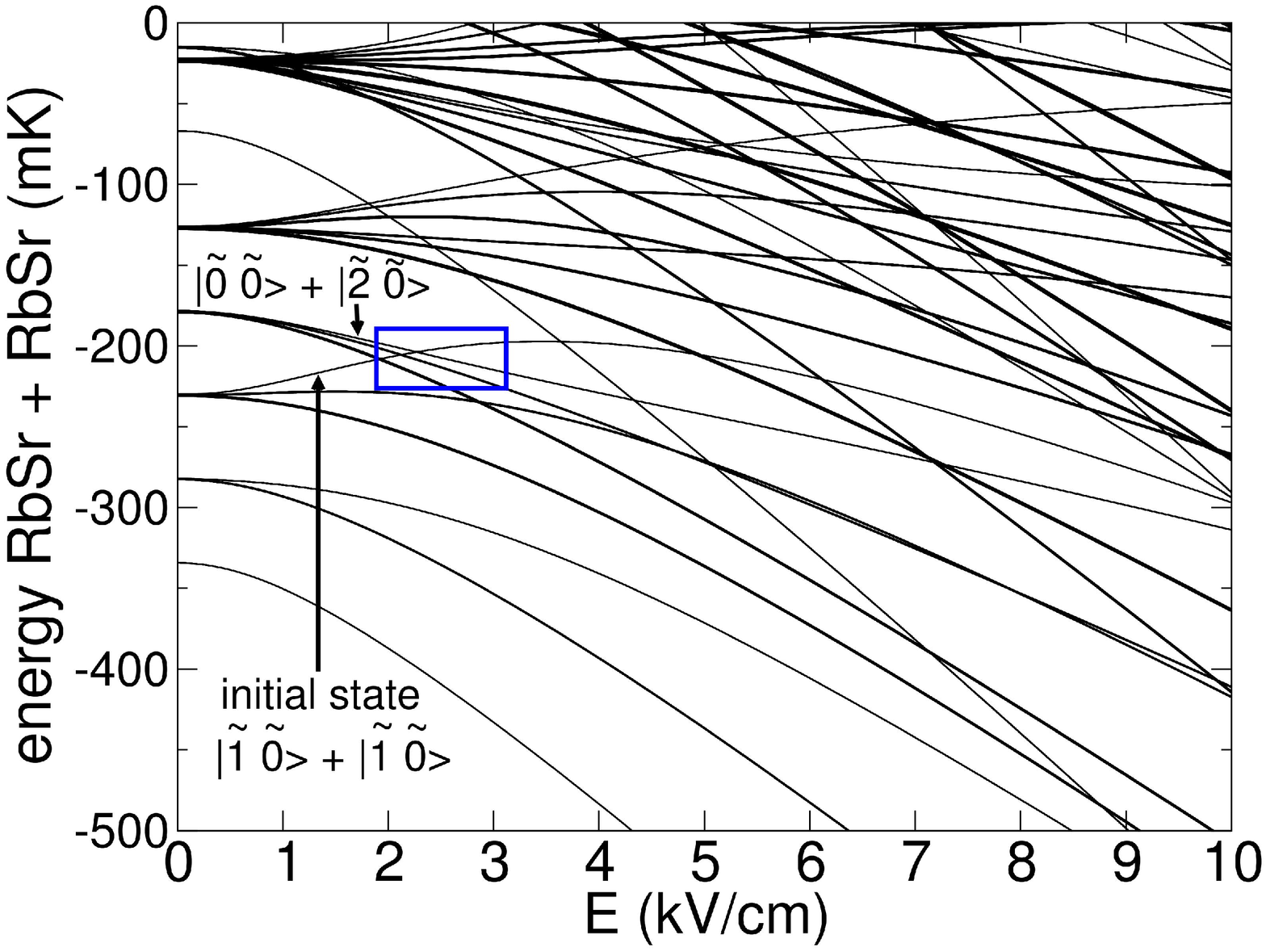}
\end{center}
\caption{(Color online) Energy of the combined RbSr + RbSr molecular states. 
The initial colliding state 
is indicated with an arrow as well as the ``crossing'' state, at the crossing field 
$E^* = 2.27$~kV/cm indicated inside the blue box. The combined molecular states 
are denoted $|\tilde{n}_1,\tilde{m}_{n_1} \rangle + |\tilde{n}_2,\tilde{m}_{n_2} \rangle$.}
\label{NRG-2PLE-FIG}
\end{figure}

To get the energy of the molecules we diagonalize $H_{\rm mol}$ in Eq.~\eqref{HAM} in the total uncoupled basis set, using $n=0-3$ and the other values given in Table \ref{TAB1}.
The resulting energy of the RbSr molecule is displayed as a function of the electric field in 
Fig.~\ref{NRG-1PLE-FIG}. To a good approximation, each state can be labeled by the rotational quantum number $n$ to which it correlates at zero field; we denote this number as ${\tilde n}$ as a reminder that $n$ is not strictly conserved.  Likewise, even though $m_n$ is not strictly a good quantum number, it is useful to denote states by the approximate value ${\tilde m}_n$.
Note that the hyperfine structure is too small to be seen in the figure.
As explained in the Introduction, likely candidates for molecules that exhibit dipolar shielding would be those in the upper state of a parity doublet.  To this end, we choose the first rotationally excited state $|{\tilde n}, {\tilde m}_n \rangle = |\tilde{1} \tilde{0} \rangle$ as indicated in the figure. This state will be opposite in parity to the nearby states $| \tilde{0} \tilde{0} \rangle$ and $| \tilde{2} \tilde{0} \rangle$.

Figure~\ref{NRG-2PLE-FIG} shows the energies of pairs of RbSr molecules, as a function of electric field. These energies define the collision thresholds. For purposes of shielding, the useful case is where the energy of the pair $|\tilde{1} \tilde{0}\rangle$ + $|\tilde{1} \tilde{0} \rangle$ just exceeds the energy of the pair $|\tilde{2} \tilde{0} \rangle$ + $|\tilde{0} \tilde{0} \rangle$, belonging to different parity states~\cite{Avdeenkov_PRA_73_022707_2006,Wang_NJP_17_035015_2015}. The corresponding electric field range, just above a critical field $E^* =2.27$ kV/cm for RbSr, is indicated by a blue box in Fig.~\ref{NRG-2PLE-FIG}.  It is in this electric field range, where the initial state of interest lies {\it above} states of opposite parity, that we expect repulsive long-range interactions.

The collisional formalism has been described in detail elsewhere~\cite{Avdeenkov_PRA_73_022707_2006,Wang_NJP_17_035015_2015}. 
The time-independent Schr{\"o}dinger equation leads to a set of coupled differential equations 
which is solved using the propagation of the log-derivative 
method~\cite{Johnson_JCP_13_445_1973,Manolopoulos_JCP_85_6425_1986}. The asymptotic matching of the wave function to free-particle wave functions  at 
large radial molecule-molecule separation $r$ provides the cross sections and rate coefficients as 
a function of the collision energy $E_c$ and the electric field $E$.
The basis set employed to express the Schr{\"o}dinger equation into a set of differential 
equations is made of the dressed states 
$|\tilde{n}_1,\tilde{m}_{n_1} \rangle$ of molecule 1
(short for $|\tilde{n}_1,\tilde{m}_{n_1},\tilde{s}_1,\tilde{m}_{s_1},\tilde{i}_1,\tilde{m}_{i_1} \rangle$),
the dressed states $|\tilde{n}_2,\tilde{m}_{n_2} \rangle$ of molecule 2
(short for $|\tilde{n}_2,\tilde{m}_{n_2},\tilde{s}_2,\tilde{m}_{s_2},\tilde{i}_2,\tilde{m}_{i_2} \rangle$) and partial waves $|l,m_l\rangle$.
The total angular momentum projection quantum number for the 
collision $M = m_{f_1} + m_{f_2} + m_l$
is conserved during the collision.

The potential energy consists of an isotropic 
van der Waals interaction and the dipole-dipole
interaction between the two molecules~\cite{Avdeenkov_PRA_73_022707_2006,Wang_NJP_17_035015_2015}.
We use a value of $C_6=-15253$~E$_h$.a$_0^6$ for 
the electronic van der Waals coefficient
between two RbSr molecules~\cite{Zuchowski_2013}.
The value of this coefficient has not been calculated for the other molecules SrF, BaF, YO,  
therefore we use a fixed value of $C_6=-10000$~E$_h$.a$_0^6$ which might
be a generous upper limit for molecules containing F or O atoms.
In any case, the van der Waals interaction plays a significant role only
at vanishing electric fields and we believe that this chosen value
will not affect the results at higher fields, where dipolar forces dominate.  

In general, these $^2\Sigma$ molecules would interact via both singlet and triplet potentials at short range.  We do not make this distinction in the present treatment, as we are concerned primarily with the effects of long-range shielding, at intermolecular separations where the singlet and triplet potentials are degenerate.  Possible triplet-to-singlet transitions for spin-polarized molecules are subsumed within the schematic treatment of short-range physics.  Namely,
in all calculations, it is assumed that the molecules are lost with unit probability
when they meet at short-range ($r = 10$~a$_0$). The loss mechanism is either due to chemical reactions or, if not reactive, 
due to long-lived four-body complexes destroyed by other colliding molecules~\cite{Mayle_PRA_87_012709_2013}. 
This loss is modeled by applying absorbing boundary conditions at the radius $r$
where the propagation of the log-derivative matrix is started~\cite{Wang_NJP_17_035015_2015}.

We restrict our attention to collisions of identical bosons, since those would have the highest reaction rate for s-wave scattering at ultralow temperature, but similar results would be found for identical fermions~\cite{Avdeenkov_PRA_73_022707_2006,Wang_NJP_17_035015_2015}. 
For each calculation, we compute the elastic $\beta_{el}$, inelastic $\beta_{in}$ and reactive $\beta_{re}$ rate coefficients. $\beta_{qu} = \beta_{in} + \beta_{re}$ 
is the quenching rate coefficient corresponding to the processes that lead to loss of molecules in an ultracold trap. The figure of merit for evaporative cooling is the ratio $\gamma = \beta_{el}/\beta_{qu}$ of elastic to quenching rates.  This ratio should, as a rule of thumb, exceed about 100 for evaporation to be at all successful.

\section{Results for shielding RbSr molecules}

\subsection{Complete Spin Structure}

In the present section, we use the full long range Hamiltonian, including the fine and hyperfine structure. However, as a first approximation, we restrict the number of partial waves to two, using only $l=0,2$ to reduce the numerical cost. We set $M=0$ which is the lowest projection of the total angular momentum, appropriate for low collision energies of indistinguishable bosons. 
We note already how big the numerical cost is when we include the fine and hyperfine structure. Without the condition of constant $M$, if we use $n=0-3$ as mentioned above, it corresponds to 16 rotational states. For RbSr, we have 2 states for $s=1/2$ and 4 states for $i=3/2$. One dressed molecular state corresponds then to a linear combination 
of $16 \times 4 \times 2 = 128$ bare molecular states. The number of combined molecular states is then $128  \times 129 / 2 = 8256$ and the number of partial waves components is 6. Then the total number of channels amounts to 49536. This shows how intensive a full scattering calculation becomes. The inclusion of higher partial waves will be discussed in the next section.

\begin{figure} [h]
\begin{center}
\includegraphics*[width=6.7cm,keepaspectratio=true,angle=-90]{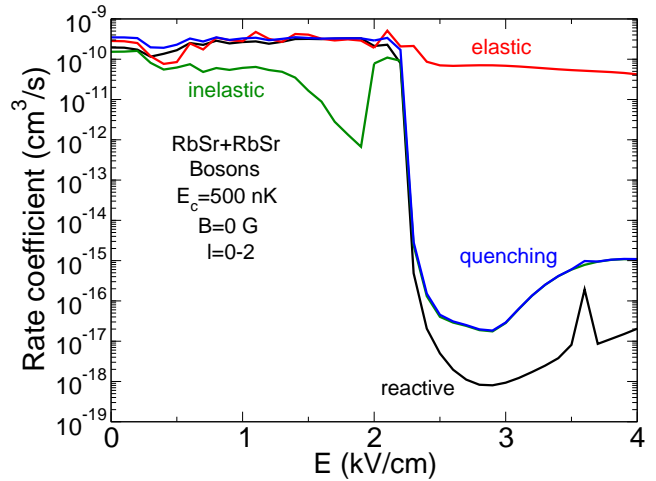}
\end{center}
\caption{(Color online) Elastic (red), inelastic (green), reactive (black), quenching (blue) rate coefficients as a function of the electric field for a fixed collisions energy of $E_c = 500$~nK,
for bosonic $^{87}$Rb$^{84}$Sr + $^{87}$Rb$^{84}$Sr collisions initially in the state indicated in Fig.~\ref{NRG-2PLE-FIG}.}
\label{RATE-EFIELD-FIG}
\end{figure}

\begin{figure} [h]
\begin{center}
\includegraphics*[width=6.7cm,keepaspectratio=true,angle=-90]{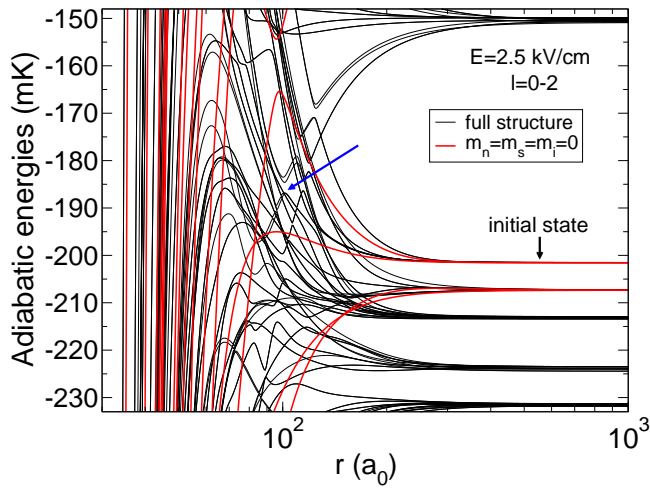}
\end{center}
\caption{(Color online) Adiabatic energies as a function of the molecule-molecule separation $r$ at a fixed field 
of $E=2.5$~kV/cm for $l=0-2$. The black curves represent a full calculation including all 
rotational states as well as the fine and hyperfine structure. 
The blue arrow indicates where the combined molecular state above the initial state 
starts to play a role. The red curves correspond to a 
simplified calculation involving $m_{n_i}=m_{s_i}=m_{i_i}=0$, $i=1,2$.
To compare with the full structure calculation,
all the red curves have been translated so that the two combined molecular states 
have the same threshold energies at large $r$.}
\label{SPAG-FIG}
\end{figure}

We present in Fig.~\ref{RATE-EFIELD-FIG} the resulting rate coefficients for the bosonic collision $^{87}$Rb$^{84}$Sr + $^{87}$Rb$^{84}$Sr. 
From $E = 0$ to 2 kV/cm, the rate coefficients are quite high, taking a value $\sim10^{-10}$ cm$^3$/s typical for ultracold reactive scattering~\cite{Quemener_PRA_84_062703_2011}.
But at the crossing electric field of $E^* \approx 2.27$ kV/cm, there is a sudden and substantial drop in the inelastic and reactive rates, hence in the quenching rates, while the elastic rate coefficients change only slightly. 

The sudden drop in chemical reactivity above the critical electric field occurs for the same reason as described in Refs.~\cite{Avdeenkov_PRA_73_022707_2006,Wang_NJP_17_035015_2015}.  The effect is illustrated by the adiabatic potential energy curves shown in figure~\ref{SPAG-FIG}, constructed for an electric field value $E = 2.5$ kV/cm.  At this field, because the entrance channel is slightly higher in energy than channels with opposite parity, its corresponding adiabatic curves are repulsive, hence preventing molecules from accessing the short range physics.
This repulsive adiabatic potential ultimately turns over to become attractive, but only at smaller $r$ and due to avoided crossings with a higher potential, as indicated by the blue arrow in the figure. This turnover corresponds to an energy  
of $\sim 15$~mK from the initial threshold. 
Thus for low collision energies $E_c \le 1$~mK, the curves remain repulsive
for the incident channel, and we can understand qualitatively 
the suppression seen in Fig.~\ref{RATE-EFIELD-FIG}. The height of this potential suggests that the dipolar suppression will be quite effective at MOT temperatures.

Therefore, despite the additional fine and hyperfine structure
of the $^2\Sigma$ molecules, the suppression mechanism described in Ref.~\cite{Avdeenkov_PRA_73_022707_2006,Wang_NJP_17_035015_2015} for $^1\Sigma$ molecules
still works well here. In fact, the region of suppression of collisions extends to an even larger electric field range for RbSr than for KRb (compare Fig.~4 of Ref.\cite{Wang_NJP_17_035015_2015}).  This is because the additional spin states provide a greater number of opposite parity states to repel the adiabatic curve upward in $^2\Sigma$ molecules as compared to $^1\Sigma$ molecules. 
The shielding occurs at sufficiently long range that spin-changing mechanisms that drive the colliding partners from the triplet potential to the singlet potential are likely also suppressed.  In any event, our calculations assume unit probability of reaction at small $r$, which should accommodate any such transfer to the singlet potential and subsequent chemical loss. Should this prove not be the case, it remains possible to suppress the triplet-to-singlet transition by means of a magnetic field~\cite{Hummon_PRL_106_053201_2011}.

\subsection{Spin degrees of freedom as spectators}

The mechanism of dipolar shielding relies on creating the repulsive adiabatic potential depicted in Fig.~\ref{SPAG-FIG}. This in turn depends primarily on two factors: i) the dipole-dipole interaction, and ii) the proximity of molecule-pair states of opposite parity to the initial states, at a slightly lower energy than the initial states themselves.  Neither of these factors relies heavily on the spin or nuclear spin of the molecules, except perhaps to alter details.  This circumstance should allow for some simplifications in the calculation, which we explore here.

First we note that, for $E$ just above the crossing field $E^*$, the two most significant 
combined molecular states for the shielding mechanism, 
the initial state $|\tilde{1} \tilde{0} \rangle |\tilde{1} \tilde{0} \rangle$ and the crossing state $| \tilde{2} \tilde{0} \rangle |\tilde{0} \tilde{0} \rangle$, are nearly degenerate 
and then well isolated from all the others.
All of these molecular states have $m_{f_i} = m_{n_i} + m_{s_i} + m_{i_i} = 0$ with $i=1,2$.
This suggests a first simplification that consists of retaining only those scattering channels 
with total rotational quantum number $m_{f_1}=m_{f_2}=0$.

Second, we note that the $^2\Sigma$ molecules we are considering are very well described by Hund's case b), as can be seen by their small spin-rotation coupling constants in Table~\ref{TAB1}.
In an electric field, different rotational quantum numbers are mixed and 
the rotational structure is strongly affected. The diatomic 
molecular frame precesses around the electric field. However
the electronic and nuclear spins are nearly decoupled from the rotation
and are not significantly affected by the electric field.
They should therefore act primarily as spectators to the shielding mechanism.
The second simplification therefore neglects the contribution of the electronic 
and nuclear spin quantum numbers, namely $s,m_s,i,m_i=0$, leading to $V_{ns} = V_{si}=0$. 
This implies, along with the first simplification, that all the individual 
projections $m_{n_i}=m_{s_i}=m_{i_i}=0$ with $i=1,2$.

Upon making these simplifications, the number of combined molecular states in the calculation is substantially reduced.
This is seen in Fig.~\ref{SPAG-FIG} where the red curves correspond
to the adiabatic energies using the second simplification ($m_{n_i}=m_{s_i}=m_{i_i}=0$, $i=1,2$).
The energies have been translated so that the threshold 
energies of the two main states 
match the ones of the full calculation. The other combined molecular states 
that were present in the full calculation are now absent. 
But the repulsive curves 
for the initial state are properly described, at least for 
collision energies $E_c \le 1$~mK.
The essential features of the shielding therefore remain intact.

\begin{figure} [t]
\begin{center}
\includegraphics*[width=6.7cm,keepaspectratio=true,angle=-90]{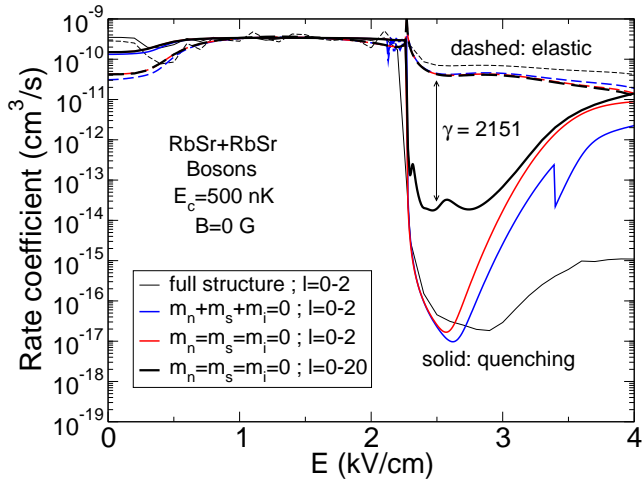}
\end{center}
\caption{(Color online) Quenching (solid lines) and elastic (dashed lines) 
rate coefficients of $^{87}$Rb$^{84}$Sr + $^{87}$Rb$^{84}$Sr collisions
as a function of the electric field for the partial waves $l=0-2$
for a fixed collisions energy of $E_c = 500$~nK. Two simplifications are used: 
$m_{f_i} = m_{n_i} + m_{s_i} + m_{i_i} = 0$ (blue lines) and $m_{n_i}=m_{s_i}=m_{i_i}=0$ (red curves). The thin black lines 
recall the same calculation with the full rotational and spin structure from 
Fig.~\ref{RATE-EFIELD-FIG}. Finally, the thick black lines is the converged result with the 
partial waves $l=0-20$ for the $m_{n_i}=m_{s_i}=m_{i_i}=0$ simplification.}
\label{RATE-EFIELD-APROX-FIG}
\end{figure}

The corresponding rate coefficients for the two simplifications are shown 
in Fig.~\ref{RATE-EFIELD-APROX-FIG}, 
in blue ($m_{f_i} = m_{n_i} + m_{s_i} + m_{i_i} = 0$) and in red ($m_{n_i}=m_{s_i}=m_{i_i}=0$) curves. 
Both approximations adequately describe
the rate constant near the crossing field despite 
the substantial simplifications.
This implies that the shielding mechanism
is mainly explained by the $m_n=0$ rotational 
structure of the molecules of the 
two combined molecular states nearly degenerate,
and not the spin structure.
Above $E \sim 2.7$ kV/cm, 
the approximations are no longer adequate, since repulsion from the other, neglected, 
spin states is required for further suppression.
The shielding is still present but cannot be explained solely by the two combined
molecular states. It is therefore not valid to use the present simplifications too far from the crossing field.

\begin{figure} [t]
\begin{center}
\includegraphics*[width=6.7cm,keepaspectratio=true,angle=-90]{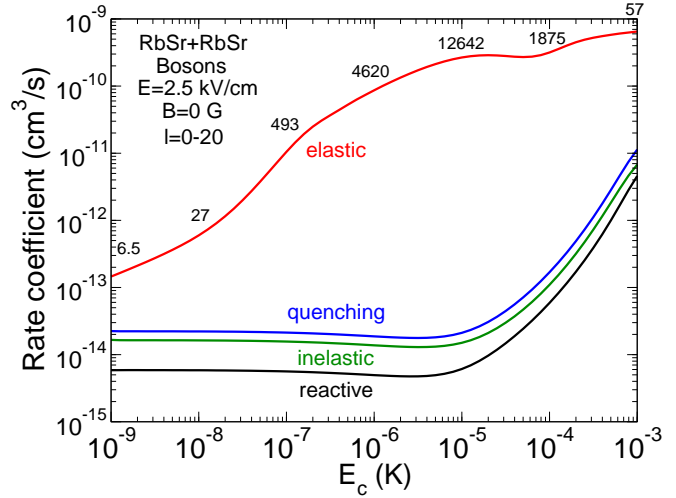}
\end{center}
\caption{(Color online) Same as Fig.~\ref{RATE-EFIELD-APROX-FIG}
but as a function of the collision energy for a fixed electric field of $E=2.5$~kV/cm,
for the $m_{n_i}=m_{s_i}=m_{i_i}=0$ simplification. The range of partial waves is $l=0-20$ and $M=m_l=[-8,8]$. The ratio $\gamma = \beta^{el}/\beta^{qu}$ is indicated for every decade of the collision energy on top of the elastic rate curve.}
\label{RATE-ECOLL-APROX-FIG}
\end{figure}

\subsection{Role of partial waves}

We turn now to the role of partial waves.  These must be significant, since they are required to describe details of the anisotropic dipolar interaction.  
The removal of irrelevant spin degrees of freedom as discussed above implies a lower number of combined molecular states in the numerical calculation. For comparison, 
the $m_{f_i} = m_{n_i} + m_{s_i} + m_{i_i} = 0$ simplification leads to 300 combined molecular states, and the additional simplification  $m_{n_i}=m_{s_i}=m_{i_i}=0$  leads to a mere 10 states, far lower than the 8256 states required by the full structure.
Using these numerical simplifications, we are free to check the convergence of the quenching rate with increasing number of partial waves. In the following, we use the $m_{n_i}=m_{s_i}=m_{i_i}=0$ simplification.

Under these restrictions, we find that convergence requires partial waves up to $l=20$.  The converged result is shown as the thick black line in Fig.~\ref{RATE-EFIELD-APROX-FIG}.  
The inclusion of additional partial waves increases the rate coefficient 
up to three orders of magnitude compared to the one with $l=0-2$. 
This is because  the partial waves get strongly mixed by the 
dipole-dipole interaction when the two 
combined molecular states become nearly degenerate near the crossing field. 
Far from the crossing field, 
both curves with small and large number of partial waves provide the same result showing the weak 
effect of partial waves for the approximation, as the initial state get more isolated 
from the others.

We finally note that the simplified models do not reproduce the resonant structure seen at fields below
$E \le 2$~kV/cm.
The blue and red curves in Fig.~\ref{RATE-EFIELD-APROX-FIG} from approximations
are smooth in contrast with the thin black curves of the full 
calculation, where we see the presence of scattering resonances in the rates 
(also clearly seen in Fig.~\ref{RATE-EFIELD-FIG}).
Usually, scattering resonances from short range contributions should be totally washed out due to 
the full absorbing condition at short range~\cite{Idziaszek_PRA_82_020703_2010,Wang_NJP_17_035015_2015}. Those resonances 
should then come from the long range part only, where the complex network of  
adiabatic energies (see Fig.~\ref{SPAG-FIG}) can provide an additional scattering phase shift due 
to the interplay of the fine and hyperfine structure. Despite 
the presence of the absorbing potential at short-range, this additional phase shift
leads to a scattering resonance structure in the rate for the full calculation. 
When we drastically simplify the network of
curves  in Fig.~\ref{SPAG-FIG} using the $m_{n_i}=m_{s_i}=m_{i_i}=0$ approximation, 
the adiabatic energies in red are far less complicated at long range and provide 
no additional scattering phase shift. Then the short-range absorbing potential entirely washes out 
resonances in the rate coefficients, 
as we can see for the results of the two simplifications.

\section{Prospects for evaporative cooling}

\subsection{Energy dependence of rates for RbSr}

For predicting quenching rates, we therefore use the simplifications discussed in the previous section, namely, we consider only channels with $m_{n_i}=m_{s_i}=m_{i_i}=0$, and we include partial waves up to $l=20$ to account for the dipole-dipole interaction. 
Using this model, we compute the rate 
coefficients as a function of the collision energy. 
To converge the results, mostly at the highest energies around 1 mK and mostly for the elastic rates, we included 
the components $M=m_l=[-8,8]$. At $E_c = 1$~mK, this enables a convergence of $\sim 4 \%$ for the elastic rates and less than $0.01 \%$ for the quenching rates compared to a calculation with the components $M=m_l=[-7,7]$.
This calculation is presented, for RbSr molecules, in 
Fig.~\ref{RATE-ECOLL-APROX-FIG} for a fixed electric field of $E = 2.5$~kV/cm 
(slightly above the crossing field), for a range of collision energies 1 nK to 1 mK.
From $\sim 40$ nK to $\sim 0.6$ mK the ratio of the elastic over the quenching rates $\gamma \sim [100 - 10000]$
looks favorable for evaporative cooling purposes where
strong elastic processes are needed.
We recall here that this result is for bosonic molecules starting in $\tilde{n}=1$ states
for unconfined, free-space collisions. In contrast 
for ground state molecules in $\tilde{n}=0$ states, 
$\gamma < 1$~\cite{Julienne_PCCP_13_19114_2011}. 
The ratio $\gamma$ decreases as the collision energy increases because the molecules 
have a greater chance to overcome the repulsive curves in Fig.~\ref{SPAG-FIG} and then
access the short-range region or undergo an inelastic rotational transition.
The ratio also decreases when the collision energy decreases due to the 
Wigner laws $\gamma = \beta^{el}/\beta^{qu} \sim \sqrt{E_c}$ as $E_c \to 0$.
Therefore there is just a limited range of collision energy (depending on the system)
where the ratio should be favorable for evaporative cooling. Luckily, this range coincides neatly with the range that needs to be spanned to get from cold MOT temperatures to truly ultracold temperatures in the single partial wave regime.

\begin{figure} [t]
\begin{center}
\includegraphics*[width=6.7cm,keepaspectratio=true,angle=-90]{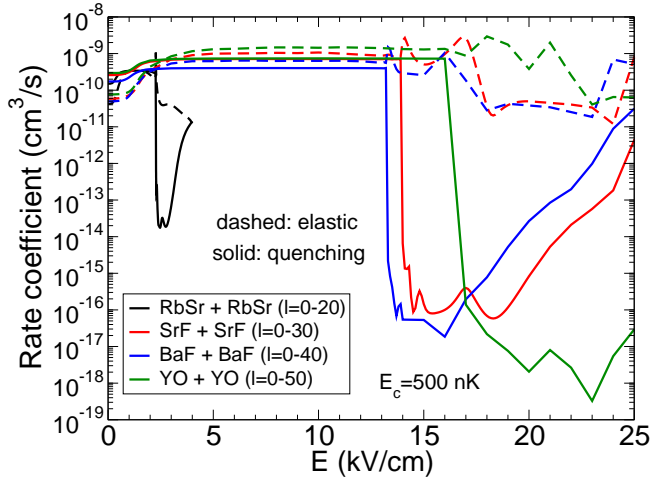}
\end{center}
\caption{(Color online) Quenching (solid lines) and elastic (dashed lines) rate coefficients of $^{84}$Sr$^{19}$F + $^{84}$Sr$^{19}$F (red), 
$^{138}$Ba$^{19}$F + $^{138}$Ba$^{19}$F (blue)
and $^{89}$Y$^{16}$O + $^{89}$Y$^{16}$O (green) collisions 
as a function of the electric field, for a fixed collisions energy of $E_c = 500$~nK. 
The range of partial waves is respectively 
$l=0-30$, $l=0-40$, $l=0-50$.
The curve from Fig.~\ref{RATE-EFIELD-APROX-FIG} for RbSr has been added for comparison.}
\label{RATE-EFIELD-APROX-OTHER-FIG}
\end{figure}

\begin{figure} [t]
\begin{center}
\includegraphics*[width=6.7cm,keepaspectratio=true,angle=-90]{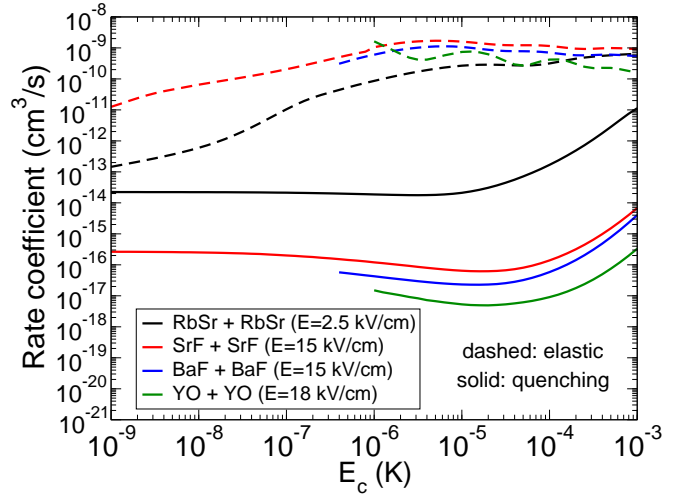}
\end{center}
\caption{(Color online) Same as Fig.~\ref{RATE-EFIELD-APROX-OTHER-FIG} 
but as a function of the collision energy 
for a fixed electric field of $E=15$~kV/cm for the SrF and BaF molecules, 
and $E=18$~kV/cm for the YO molecules.
The curve from Fig.~\ref{RATE-ECOLL-APROX-FIG} for RbSr has been added for comparison.}
\label{RATE-ECOLL-APROX-OTHER-FIG}
\end{figure}

The experimental quest to produce molecules in their absolute ground state, 
including the ground rotational state, continues. But loss processes are quite strong compared to elastic 
ones~\cite{Quemener_PRA_83_012705_2011} for those molecules, especially for bosons.
By simply changing the initial rotational state of the molecules by a single quantum,
from the ground rotational state to the first excited rotational state, quenching can be suppressed.
This is a promising result for current ultracold experiments of bosonic RbSr molecules~\cite{Pasquiou_PRA_88_023601_2013}.
A calculation involving the time evolution of a molecular cloud of RbSr will be needed in order to determine to what temperature the gas can be cooled to without suffering from too much molecular loss. This is beyond the scope of this paper and left for further studies.
Another interesting idea suggested by these results
would be to produce the RbSr molecules
directly in the $\tilde{n}=1$ rotational state at $E=2.5$~kV/cm
instead of the ground state $\tilde{n}=0$ at zero electric field.
In such way the molecules will be protected from
loss directly during their STIRAP~\cite{Ni_S_322_231_2008,Danzl_S_321_1062_2008} formation.
Finally in such experiments, the magnetic field used for the formation of the 
Feshbach molecules is often kept on and can 
modify substantially the landscape of the curves 
seen in Fig.~\ref{NRG-1PLE-FIG}, \ref{NRG-2PLE-FIG} and \ref{SPAG-FIG}.
Therefore a magnetic field can also play a role in the rates and the ratio $\gamma$,
a task which is also left for future studies.

\subsection{Prospects for other molecules: SrF, BaF and YO}

Based on the same $m_{n_i}=m_{s_i}=m_{i_i}=0$ approximation, we can also investigate other molecules of current experimental interest.  Here we focus on
bosonic $^{84}$Sr$^{19}$F + $^{84}$Sr$^{19}$F, $^{138}$Ba$^{19}$F + $^{138}$Ba$^{19}$F and $^{89}$Y$^{16}$O + $^{89}$Y$^{16}$O collisions. 

The rates as a function of the electric field are reported in Fig.~\ref{RATE-EFIELD-APROX-OTHER-FIG} for a fixed collision energy of $E_c = 500$~nK.
We see a more pronounced but similar shielding mechanism for all three systems compared to the alkali-alkaline earth RbSr molecule. This is due to the twice to thrice bigger 
electric dipole moment of the molecules compared to RbSr, hence increasing 
the dipole-dipole couplings between the initial and the crossing combined molecular states
and then increasing the repulsive curves that shield the collision.
Correspondingly, the number of partial waves required to converge the results is greater than for RbSr molecules
($l=0-20$). For the SrF, BaF, YO system we need respectively $l=0-30$, $l=0-40$, $l=0-50$ partial waves. 
We used the components $M=m_l=[-4,4]$ for SrF and BaF.
At $E_c = 1$~mK, the elastic rates are converged to $16 \%$ for SrF and $15 \%$ for BaF, while the quenching rates to $2 \%$ for SrF and $1.5 \%$ for BaF. 
Note that the elastic rates, even though not as converged as the quenching rates, can only be bigger than the presented values here at high collision energies. For YO we performed the calculation only for $M=m_l=0$, for which we believe that the values of the elastic and quenching rates are converged within a factor of 2 (at the highest collision energies) but that their ratio $\gamma$ is less affected so that the main conclusion of the shielding mechanism is unchanged, as we will see.

The rates as a function of the collision energy for fixed electric fields of 
$E=15$~kV/cm (SrF, BaF) and $E=18$~kV/cm (YO) where the shielding takes place
are shown in Fig.~\ref{RATE-ECOLL-APROX-OTHER-FIG}. The three systems offer a remarkably
high ratio $\gamma$, where $\gamma > 10000$ 
in the range $E_c =[1 \text{nK} - 1 \text{mK}]$ of collision energy for SrF and 
in the range $E_c =[1 \mu\text{K} - 1 \text{mK}]$ for BaF and YO. 
It is seen that the particular values of the elastic and quenching rates don't matter much since the quenching rates are highly suppressed compared to the elastic ones.
Evaporative cooling in free space conducted in the appropriate electric field
could be therefore a very optimistic method for cooling further down those systems to reach quantum degeneracy.

\section{Conclusion and perspectives}

Shielding $^2\Sigma$ molecules against collisional losses 
over a wide range of collision energies is possible
if the molecules are addressed in their first rotational quantum state 
at a particular shielding electric field.
This is due to the strong coupling with a nearby, almost resonant
combined molecular state 
which is brought by the electric field just below the initial colliding state.
The shielding is very effective as the permanent electric dipole moment 
of the molecules increases. This leads to a large ratio of
elastic to quenching rates  
which is certainly favourable for evaporative cooling
of dipolar gases of $^2\Sigma$ molecules.
We found that the fine and hyperfine structure of the $^2\Sigma$ molecules
can be neglected at the particular shielding electric field,
as the shielding mechanism is mainly driven by the rotational structure
of the molecules.
However the partial waves play an important role in the rate coefficients
as the two combined molecular states involved in the process
are almost degenerate and hence strongly coupled by the dipolar interaction.

Future work can imply for example the efficiency of the thermalization
provided by the elastic collisions during the evaporative cooling process, 
as well as the role of an additional magnetic field on the dynamics of the molecules.
The same shielding process can also work for molecules still in their first rotational state
but for excited vibrational state. The outcome might be different since both the 
rotational constant and the permanent dipole moment of the molecule 
depend on the vibrational quantum number~\cite{Aymar_JCP_122_204302_2005,Vexiau_JCP_142_214303_2015}.
It can be interesting to study whether the shielding can be further improved for a given system
by addressing the molecules in a higher vibrational state.
For molecules formed by a STIRAP process such as RbSr, 
it would be interesting to study whether the molecules can be formed directly
at the shielding electric field in the $\tilde{n}=1$ rotational state, 
so to protect directly the molecules during their formation.
Finally, this mechanism looks also promising for shielding three-body collisions
of dipolar molecules which might play a significant role
in dense dipolar Bose-Einstein Condensates
and might have implications for many-body physics~\cite{Petrov_PRL_112_103201_2014}. \\

\section*{Acknowledgments}

G.Q. acknowledges the financial support of the COPOMOL project (\# ANR-13-IS04-0004-01) from Agence Nationale de la Recherche, the project Attractivit\'e 2014 from Universit\'e Paris-Sud
and the project FEW2MANY (2014-0035T) from Triangle de la Physique.
J.L.B. acknowledges support from an ARO MURI, Grant No. FA9550-1-0588. 
We gratefully acknowledge discussions with Dave DeMille, Florian Schreck, and Jun Ye.
We thank Piotr $\dot{\text{Z}}$uchowski for providing us the electronic $C_6$ coefficient
between two RbSr molecules.

\bibliography{../../../BIBLIOGRAPHY/bibliography.bib}
 
\end{document}